# Accelerator Testing of the General Antiparticle Spectrometer, a Novel Approach to Indirect Dark Matter Detection

**C J Hailey[a], T Aramaki[a], W W Craig[b], L Fabris[b], F Gahbauer[a,d], J E Koglin[a], N Madden[b], K Mori[c], H T Yu[a], K P Ziock[b]**

[a]Columbia Astrophysics Laboratory, Columbia University, New York, NY
[b]Lawrence Livermore National Laboratory, Livermore, CA
[c]Canadian Institute of Theoretical Astrophysics, Toronto, Canada
[d]University of Latvia, Riga, Latvia

E-mail: chuckh@astro.columbia.edu

**Abstract**: We report on recent accelerator testing of a prototype general antiparticle spectrometer (GAPS). GAPS is a novel approach for indirect dark matter searches that exploits the antideuterons produced in neutralino-neutralino annihilations. GAPS captures these antideuterons into a target with the subsequent formation of exotic atoms. These exotic atoms decay with the emission of X-rays of precisely defined energy and a correlated pion signature from nuclear annihilation. This signature uniquely characterizes the antideuterons. Preliminary analysis of data from a prototype GAPS in an antiproton beam at the KEK accelerator in Japan has confirmed the multi-X-ray/pion star topology and indicated X-ray yields consistent with prior expectations. Moreover our success in utilizing solid rather than gas targets represents a significant simplification over our original approach and offers potential gains in sensitivity through reduced dead mass in the target area.

The general antiparticle spectrometer (GAPS) is a novel concept for detection of antimatter. As described below it is particularly well suited for antideuteron searches. Antideuterons provide an indirect signature of neutralino annihilation, as first pointed out by Donato and collaborators [1], and a direct signature of black hole evaporation [2]. The operating principles, designs and sensitivity calculations for potential balloon and satellite-based GAPS experiments have been previously reported in Mori et al. [3], but the GAPS concept has not hitherto been demonstrated in a working prototype. We report here interim progress on analyzing a GAPS prototype exposed to an antiproton beam, as well as various beams representative of cosmic backgrounds, at the KEK accelerator facility in Japan. After a discussion of the use of antimatter in dark matter and primordial black hole searches, we briefly summarize recent theoretical studies of antideuteron searches and their relevance to our ongoing experimental program. We then follow with a description of the GAPS concept, the GAPS prototype experiment and analysis and plans for continued experimental work. We note that our preliminary analysis suggests that the GAPS concept is at least as promising as our previous simulations suggested, and thus recent theoretical analyses based on reference [3] remain unaltered by the current experimental landscape.

## 1. Current Approaches to Dark Matter Detection

Many searches are underway to identify the particle nature of cold dark matter (CDM). A strong candidate is the type of weakly interacting massive particle (WIMP) that arises in supersymmetric theories (SUSY). The neutralino is the lightest supersymmetric partner (LSP) in many models, so much effort has centered on neutralino searches. Currently, the most intensive experimental activities are associated with underground dark matter searches. These searches attempt to detect neutralinos through their direct recoil interactions with target nuclei [4]. Numerous detection schemes are employed including bolometers and solid or liquid scintillators. A primary challenge is to distinguish the recoil signal of the neutralino from that of neutrons produced in muon collisions and by natural radioactivity. A recent review by Ellis et al. [5] reevaluated direct detection in light of revised estimates of the top quark mass and pion-nucleon sigma term, as well as the WMAP precision determination of the CDM density. His conclusion is that the best current experiment, CDMS II, does not probe any of the parameter space of the constrained minimal supersymmetric model (CMSSM) at the 90% confidence level. Thus third generation experiments are already being actively investigated, which aim to provide the several orders of magnitude improvement in sensitivity necessary to probe deeply into the CMSSM parameter space.

An alternate approach is to search for indirect signatures of CDM. The processing of the hadronic and leptonic debris of neutralino-neutralino annihilations leads to the generation of many indirect signatures including antiprotons, gamma-rays, positrons and neutrinos. Several experiments have detected indirect signatures suggested to be of CDM origin. A small enhancement in the cosmic positron spectrum was observed by the HEAT experiment [6,7]. This is most likely due to conventional cosmic-ray processes, but neutralino annihilation has been invoked by Kane et al. [8] and Baltz et al. [9]. Recently the diffuse 511 keV halo observed by INTEGRAL has been interpreted as $e^+e^-$ annihilation radiation produced in dark matter annihilation by Boehm et al. [10] (In this particular case the origin of the CDM would not be associated with a SUSY model though). Next generation neutrino experiments also show some prospects for detecting neutrinos from neutralino annihilation in the sun [11].

Perhaps the most intensively investigated indirect signature is that of antiprotons. The challenge is to distinguish the primary antiprotons produced by neutralino annihilation from the "background" secondary and tertiary antiprotons. Secondary antiprotons are produced in cosmic-ray (CR) collisions with protons in the interstellar medium (ISM). Tertiary antiprotons are produced in CR collisions with helium and through scattering of antiprotons during transport. The disentangling of the primary signal is best done at low energies, but the similarity in the spectral shape of the three components makes this enormously challenging [12,13]. This is made even more complicated by solar modulation and by intrinsic uncertainties in modeling the secondary and tertiary spectrum with sufficient accuracy. The problem is illustrated in figure 1a, where a recent compilation of data is shown with some theoretical models superimposed [14]. The current data at low energies are consistent with CR models and their uncertainties. Future progress requires much more accurate fits to the data. Lionetto, Morselli and Zdravkovic [15] conclude that currently no single, unified model exists which fits all the CR data on antiprotons, positrons and heavier particles as a function of energy. Use of antiprotons to constrain SUSY models in the future depends totally on the ability to model the non-primary antiprotons with much better accuracy. One then looks for small deviations of the model fits to the "background", which may be indicative of a primary antiproton signal. This is incredibly challenging. It is often overlooked that theoretical papers on CDM searches with antiprotons assume that the CR background will, at some future time, be modeled with precision well-beyond the current situation, e.g., Profumo and Ullio [16]. Nevertheless, antiprotons can be an important signature of dark matter in some SUSY models, so current experimental efforts to reduce measurement uncertainties through design of more sensitive experiments, as well as theoretical efforts to refine CR models, are necessary.

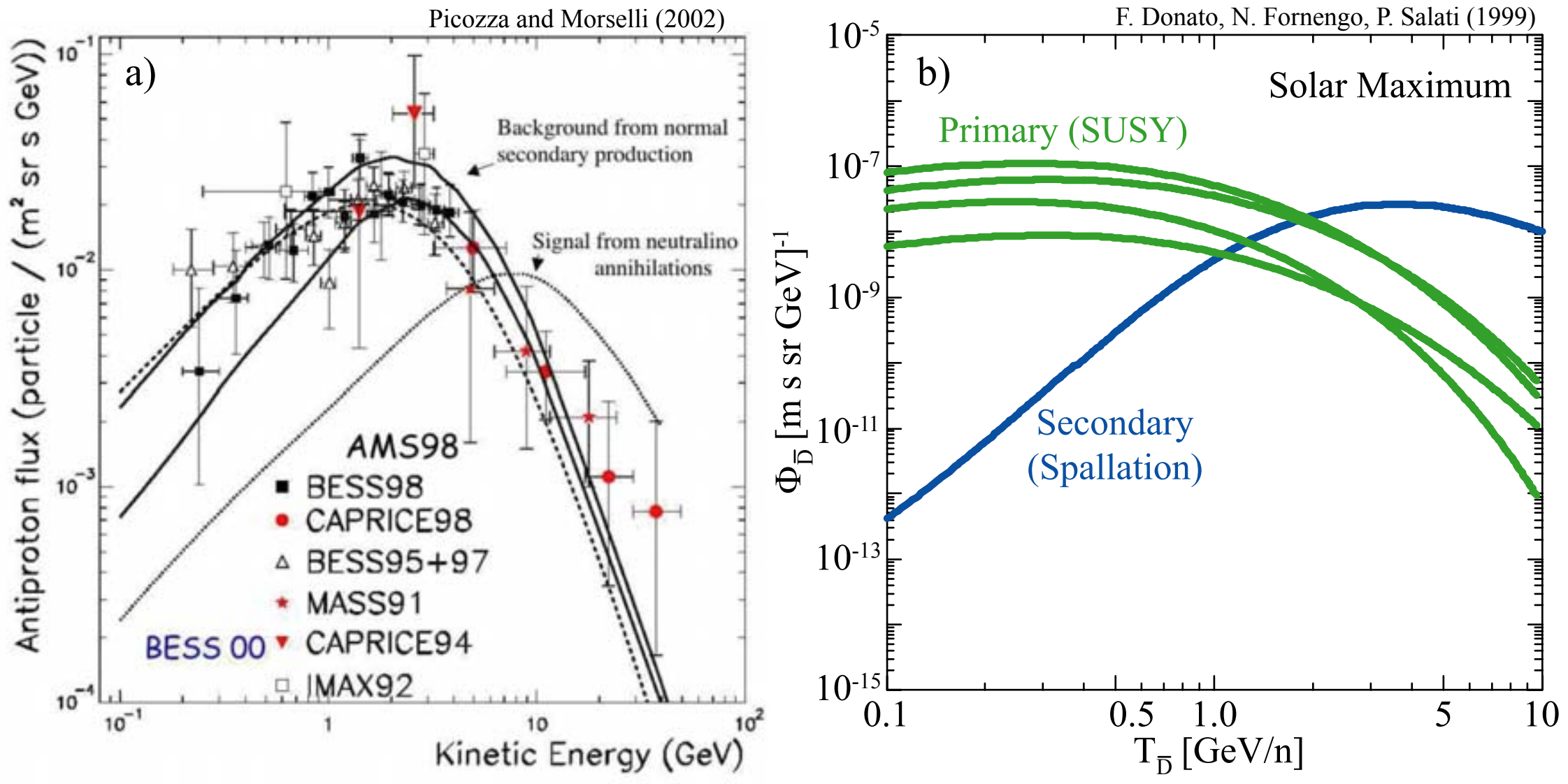

Figure 1: a) Antiproton absolute flux plotted versus kinetic energy for experimental data and theoretical predictions (replotted from Picozza and Morselli [14]). b) Solar-modulated antideuteron flux plotted versus kinetic energy per nucleon (replotted from Donato et al. [1]).

## 2. Antideuterons – A Novel Dark Matter Probe

A primary antideuteron signal in the CR presents a potential breakthrough approach for CDM searches, as pointed out by Donato et al. [1]. Secondary antideuterons, like antiprotons, are produced through the CR/ISM interaction p + A → antideuteron + X, where A is the target ISM nucleus (mainly hydrogen and helium). But the kinematic threshold for this reaction is much higher and the steep energy spectrum of CR protons means there are less with sufficient energy to produce secondary antideuterons. Thus, as envisioned by Donato et al., a low energy search for primary antideuterons below ~0.5 GeV/n is essentially free of secondary or tertiary antideuteron contamination. This is illustrated in figure 1b. Unlike primary antiproton searches, which involve fitting for small deviations from the non-primary antiproton spectrum, a low energy antideuteron search is a search – the detection of a single antideuteron can signal CDM. Figure 1b suggests that a low energy primary antideuteron search is most promising since the possibility for confusion with secondary and tertiary antideuteron background is minimized there.

Donato et al. pointed out that their results were tentative, since they did not consider in detail the effects of low energy scattering on the transport and spectrum of the secondary antideuterons. Recent papers [12] have more carefully examined the transport of secondary antideuterons at low energy. The preliminary conclusion is that the flux of tertiary (non-annihilating, inelastically scattered) antideuterons at low energies is higher than in figure 1b. However we note here that the elevated tertiary antideuteron flux predictions are not as severe a problem as for the corresponding antiproton searches. For the ultra-long duration balloon experiment we envisage, the primary antideuteron detection limit is >~ 20 times higher than the predicted tertiary antideuteron background. Detection of antideuterons in this case means primary antideuterons. Only on the very largest satellite experiments, where sensitivities and observation times are at their maximum, are primary and non-primary antideuterons likely to be confused. However this happens near the bottom of most SUSY parameter spaces. Even this qualification depends on assumptions about the halo model and clumpiness in the halo distribution, each of which sensitively affects primary antideuteron production.

## 3. Direct Comparisons of Antideuterons with Other Approaches

One of the challenges of evaluating approaches to CDM detection is that experimental sensitivities are directly coupled to model dependencies. So comparisons between various detection schemes are only meaningful when evaluated within the context of specific models. Recently a comprehensive analysis of direct versus indirect detection in minimal supergravity (mSUGRA) was presented by Edsjo, Schelke and Ullio [17]. mSUGRA is the easiest and most popular scheme for implementing SUSY breaking. The unequivocal conclusion of this analysis is that antideuterons and direct detection are the best means to search for the neutralino and constrain the mSUGRA parameter space. In some regions of parameter space the antideuteron is the only viable search approach, and it maintains sensitivity to very high neutralino mass. In other parameter regions antideuterons are competitive with direct detection and with indirect searches using neutrinos produced in neutralino annihilations in the sun.

There are more complicated SUSY models which relax one or more of the technical assumptions of the simple mSUGRA models or CMSSM. Recent papers have emphasized the power of antideuterons in these models as well. An example is the recent work of Profumo and Ullio [16] which considered a model with WIMP relic abundance enhancement in three benchmark scenarios. Using published sensitivity curves for a satellite-based GAPS experiment [3] these authors concluded that the neutralino could be detected in all three of their benchmarks (mSUGRA funnel, non-universal gaugino (NUGM) and minimal anomaly-mediated SUSY breaking (AMSB)). Although they did not mention it, even a balloon-based GAPS could detect the neutralino in the first benchmark and depending on the assumed halo CDM distribution, with modest density enhancements in the latter two benchmarks as well. Masiero, Profumo and Ullio [18] has also considered antideuteron constraints on a split-SUSY model. The conclusion is that antideuterons are crucial for neutralino searches in part of this parameter space, antiprotons are best in another part and accelerator-based work will probe none of the parameter space (because of the heaviness of squarks in these models). Within the CMSSM model there is considerable complementarity between underground dark matter searches and antideuteron searches. In fact, even an ultra-long duration balloon GAPS can cover regions of the CMSSM space which are inaccessible to third generation direct detection experiments such as XENON or GENIUS. This has been considered in some detail elsewhere [3].

For most of the SUSY models under current investigation, antideuterons are a powerful and versatile means to hunt for dark matter. In several of the references described above, there are parts of parameter space where antiprotons provide very strong dark matter signatures. However, as noted above, this will require extremely delicate extrication of the signal from the secondary and tertiary antiproton background. At any rate, antiprotons would be detected by a GAPS experiment as a byproduct of an antideuteron search, with sensitivities more than an order of magnitude higher than previously obtained [3].

## 4. Antideuterons and Primordial Black Holes

There are a variety of potential probes to search for primordial black holes (PBH): gravitational waves, gamma-rays, UHE particles, neutrinos. A recent theoretical analysis by Barrau et al. [2] indicated that antideuterons will ultimately provide the most stringent bounds on PBH density, orders of magnitude better than antiprotons and gamma-rays While antiprotons are emitted by PBH, conventional CR models already give good fits to cosmic antiproton spectra. However, we note that a small excess of low-energy antiprotons in the BESS data has been fit with a PBH model [19]. While not yet competitive with antiprotons, it is notable that the first upper limit on PBH evaporation using antideuterons has recently been presented by the BESS team [20]. Barrau et al. analyzed the ability of both AMS/ISS and a satellite GAPS to constrain PBH density. They parameterized their results in terms of both the poorly constrained diffusive halo size and the coalescence momentum (a parameter relevant to the formation of antideuterons in hadron showers). AMS/ISS would be able to improve on current PBH density limits by ~100 times and GAPS by ~1000 times. The superior GAPS limit results from higher effective grasp (area-solid angle product) and its operation at lower energies where tertiary antideuterons are less of a problem. As we will discuss in a subsequent paper this conclusion is only slightly weakened when the results of [12] on low energy antideuteron transport are taken into account.

## 5. Operating Concept of the General Antiparticle Spectrometer

The favorable spectral properties of signal and background in figure 1b come at a price; the flux of primary antideuterons is miniscule. While this flux is clearly model dependent, for experiment search times of months to years the proper order of magnitude of the grasp of an experiment is >~ 1-2 $m^2$-sr. This is to be compared with current premier magnetic spectrometer experiments such as BESS-Polar, AMS/ISS, and PAMELA, which have smaller grasps by approximately a factor of 10 [21]. In addition it is not feasible to scale up the magnetic spectrometers for next generation searches for CDM. For balloon and space-based experiments BESS and AMS likely represent the ultimate performance achievable given the respective mass limits.

GAPS was developed as a next generation antimatter detector. In Mori et al. [3] there is a detailed discussion of the atomic physics of GAPS, its design optimization and sensitivity calculations for various experiments. The interested reader is referred to this paper for a more extensive discussion. Below we describe the basic operating principles in just enough detail to elucidate the issues which must be addressed in prototype development.

A cartoon of the GAPS concept is shown in figure 2. An antiparticle first passes through a time of flight (TOF) system which measures velocity and hence energy after subsequent particle identification. It is then slowed down by dE/dx losses in a degrader block. The thickness of this block is tuned to select the sensitive energy range of the detector. The antiparticle is stopped in a target, forming an excited, exotic atom with probability of order unity. The exotic atom deexcites through both autoionizing transitions and radiation producing transitions. Through proper selection of target materials and geometry, the absorption of the antiparticle can be tailored to produce three or more well-defined X-rays in the cascade to the ground state. The target is selected such that the X-rays have energies in the 20-200 keV range so that the X-rays can escape with low losses and be efficiently detected in common X-ray detectors. After the emission of the X-rays the antiparticle annihilates in the nucleus producing a shower (star) of pions. The X-ray/pion emission takes place within nanoseconds. The fast timing coincidence between the characteristic decay X-rays of precisely known energy (dependent only on antiparticle mass and charge) and the energy deposition induced by the pion star is an extremely clean antiparticle signature. The GAPS concept is only feasible at extremely low energies (<~ 0.3 GeV/n), where particles can be ranged out with low mass degraders (essential for balloon or satellite missions where low mass is paramount). The advantage of GAPS is that for mass comparable to magnetic spectrometers it can obtain an order of magnitude more grasp. The probability of false particle identification can be reduced to 1 part in $10^{12}$ or better.

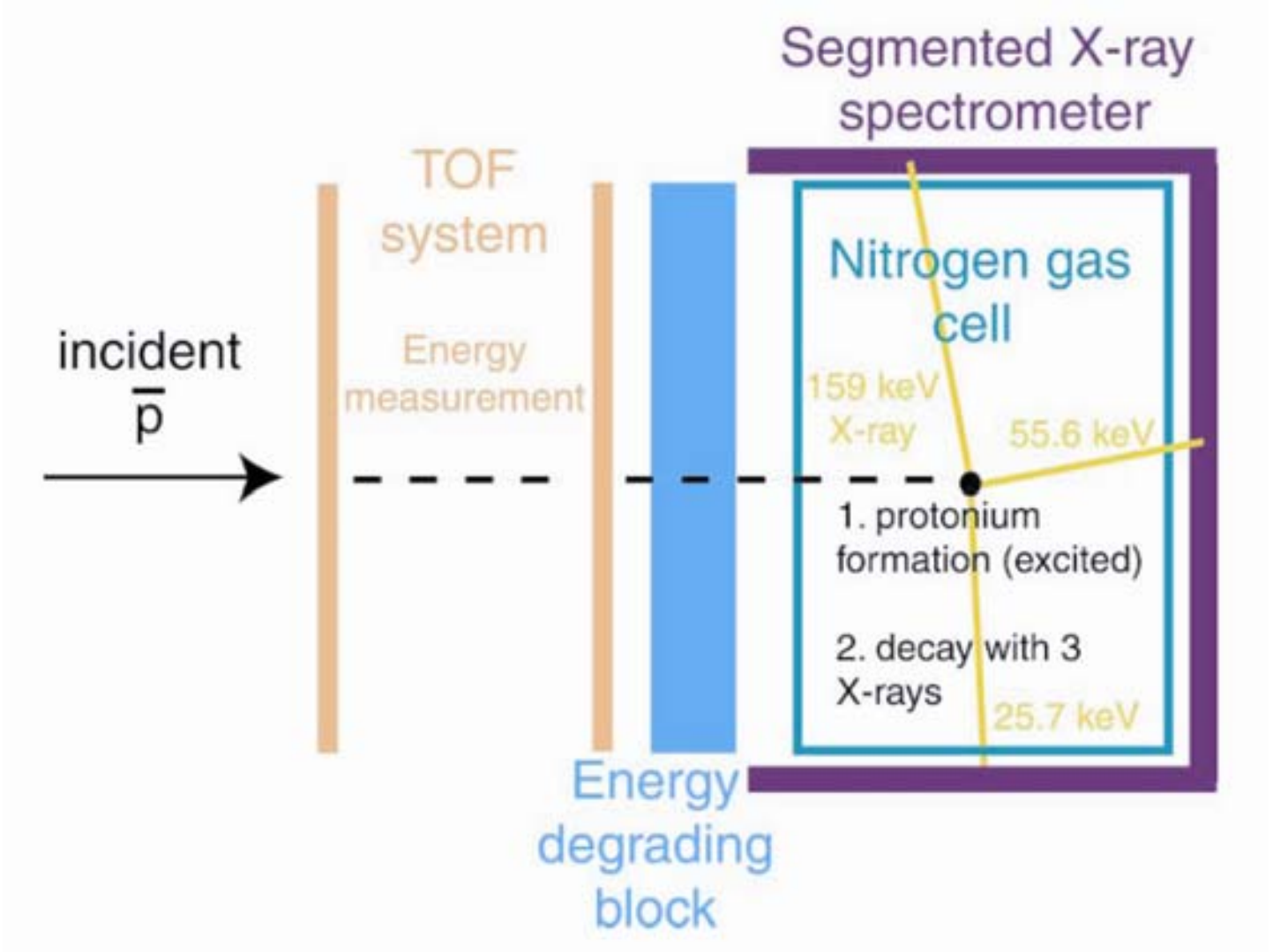

Figure 2: The operating principal of the GAPS detector using antiprotons as an example.

## 6. GAPS Crucial Design Issues

In [3] a number of designs were presented for antideuteron experiments. The designs were based on experimental and theoretical results from the literature concerning X-ray deexcitation in exotic atoms. While the results rested on a strong physics foundation, in reality no one has ever developed a detector whose specific purpose was to simultaneously measure the energies of atomic deexcitation X-rays and detect the pion star too. Moreover, while there is a wealth of data on exotic atom X-ray physics, it is often in forms not directly useful for astrophysical instrument design. For instance, X-ray transitions were measured in many gases, but not simultaneously, or they were measured for X-ray transitions whose energies were too low to be of interest in a practical detector. Consequently it was decided to build a prototype whose purpose would be to measure key parameters of interest to an astrophysical instrument designer, not an atomic physicist. To this end we discuss here some of the key design issues that were deemed necessary to address in a prototype experiment at an accelerator.

Proper target selection is the key to GAPS design. The number of simultaneously detected X-rays, together with their energies and transition yields, drive the entire design and ultimate sensitivity of the instrument. In particular our original work considered only designs with gas targets. This was because the most detailed measurements of exotic atom processes were in gases. However our prototype work, as described below, has encompassed gas, liquid and solid foil/powder/aerogel targets.

The issues in gas targets are simple. The relevant atomic physics is somewhat involved but well-understood both theoretically and experimentally [3]. High pressure (~10-30 atmospheres) is required to achieve adequate antiparticle stopping power, but must be limited in order to avoid Stark mixing which will lower the X-ray transition yields by causing non-radiative transitions. The gases nitrogen, oxygen, neon and argon are all acceptable candidates [3]. These gases are all transparent to their own ladder X-rays at appropriate design pressures. Prior to our first accelerator test (hereafter KEK04) we expanded our list to include organic gases. Higher molecular weight targets translate directly into lower requisite operating pressure. In KEK04 we operated both a nitrogen and an ethyl hexafluoride target ($C_2F_6$).

Solid targets represent an attractive option for GAPS, since they do not require gas handling systems and the ensuing dead mass. Originally we looked at liquid and solid targets. But despite materials of high X-ray yield, we could not couple out the lowest energy X-rays of interest without large self-absorption. Thus we focused on gas targets. Our interest in solids was rekindled with an investigation of low density carbon aerogel targets, which were employed at KEK04. Following this success we revisited solid targets, reviewing work carried out in the 1970s by Wiegand and Godfrey [22]. Simulations suggested that their foil target geometries, along with "wool" targets, were very attractive, providing substantial increases in X-ray detection efficiency. Thus our KEK05 run (see below) focused on foil and wool targets, as well as organic liquid targets chosen so that the X-ray transitions of interest were high enough energy to avoid self-absorption. The successful demonstration of these non-gaseous targets greatly eases the design challenge of GAPS.

In addition to confirming potential target materials, the accelerator experiments had several other goals. One goal was a measurement of the effective X-ray yield. As discussed in [3], a number of processes can interrupt the ladder deexcitation of the exotic atom, leading to missing X-rays. Antiparticle identification is predicated on identification of a specific number of X-rays of known energy. If those X-rays are missing, then an antiparticle will be miscategorized as background. Thus any inefficiency in production of the relevant ladder X-rays is an effective reduction in antiparticle detection efficiency, and thus sensitivity. A key goal of the experiments was also to gather data on event topology for both antiparticles and potential cosmic background. An antiproton event produces three or more X-rays along with a pion star of ~4-5 pions equally divided among charge states. The pion star can provide valuable added information for confirming the presence of an antiparticle. Use of the pion star topology was not considered in our original sensitivity calculations. It is equally important to understand background event topologies when analyzing rare events. Thus we took data with protons, pions and muons in order to understand details of discriminating such particle interactions in the target from antiparticle interactions. We note for thoroughness that the primary source of false antiparticle identifications is not due to particle interactions directly, but rather an accidental coincidence between a particle trigger and X-ray and beta-particles generated from cosmic-ray activation of the X-ray detector [3]. In a subsequent paper we will describe in much more detail how one analyzes real GAPS data, based on the accelerator experiments and simulations. In particular the copious antiproton flux will provide an excellent internal calibration set for verification of antideuteron detection.

For thoroughness we mention some key aspects of GAPS which are not addressed in accelerator testing. Antideuterons, like antiprotons, can annihilate as they slow down in matter prior to their capture into an exotic atom. However antideuterons have unique loss modes such as Coulomb annihilation [23] and the Oppenheimer-Phillips process [24]. As noted in Hailey et al. [25] we explicitly account for these loss modes in our sensitivity calculations, along with more common loss modes such as direct disintegration. These losses are not significant, but this is by design; the overall amount of material employed for stopping antideuterons, and thus the energy acceptance, is ultimately restricted by such effects. The lack of a slow antideuteron beam anywhere in the world necessitates the following strategy for GAPS. We verify all the basic atomic physics of GAPS with antiprotons. As described in Mori et al. [3] and references therein, the scaling of the relevant atomic physics to antideuterons is only through the particle mass. The antideuteron pion star differs from the antiproton one only in the number of emitted pions. The losses of antideuterons are handled by explicit calculation, as described above. While the new JAERI facility in Japan may ultimately have a slow antideuteron capability [26], realistically the first antideuterons detected by GAPS are likely to be in space. The presence of nuclear deexcitation gamma-rays is the only difference between antideuterons and antiprotons that cannot be explicitly accounted for by antiproton measurements or modeling. These arise in the daughter nuclei produced in the antiparticle annihilation of the nucleus of the exotic atom, as first noted by Barnes et al. [27]. As we discuss below, these gamma-rays are an unexpected benefit for background rejection.

## 7. Accelerator Testing of a GAPS Prototype: Experiment Goals

A GAPS prototype was tested at the KEK accelerator facility in Tsukuba, Japan. This followed approval at KEK of two separate proposals. One run took place in April 2004, and following an upgrade to the experiment another run took place in May 2005. The following were the goals for prototype testing and all were accomplished or are in process:

- detect simultaneous ladder X-rays and pion annihilation stars,
- characterize event topologies and compare with simulations,
- test an organic gas target of high molecular weight and one "baseline" gas as listed in [3],
- test a liquid target; test solid aerogel, powder, foil and wool targets,
- evaluate GAPS under exposure to particle background (e.g., protons, pions, muons, electrons etc.).

## 8. Experimental Setup

To create antiprotons, protons must be accelerated to many GeV in energy at particle accelerator facilities. From the interaction of the accelerated protons on a target material, particle-antiparticle pairs are sometimes created including proton-antiproton pairs. These particles are collected and transported to user research areas through secondary beamlines composed of dipole and quadrupole magnets that are used to guide and focus the particles. The secondary beamline delivers the particles in a well-defined beam envelope that can be simulated with a beamline optics program such as TRANSPORT [27]. The size and divergence of the beam in the horizontal and vertical directions as well as the spread in momentum of the particles can be calculated at any point along the beamline.

The experiment was performed using the Pi2 secondary beamline of the 8 GeV proton synchrotron at KEK. The beam is steered and focused into the experimental area by means of four quadrupole and three dipole magnets, the first of which defines the momentum. The beam structure is relatively flat with a 4 second spill repetition and a 1.5 second spill length. The characteristics of the Pi2 beamline present unique challenges that drive the design of the experiment. The Pi2 beamline is unseparated such that copious quantities of kaons, pions and electrons are transported to the experimental area along with antiprotons. Pions are the dominant background event type with four orders of magnitude higher rate than the antiprotons.

A momentum of 1 GeV/c was chosen to optimize antiproton flux (which increases steeply up to ~2 GeV/c) and losses within a degrader that is required to slow down the beam to appropriate momentum for stopping in GAPS. Most of the antiprotons either undergo direct in-flight annihilation within the degrader (producing electromagnetic showers) or are scattered directly into the GAPS detector such that only a fraction of the original antiprotons are actually stopped in the target (estimated to be only 0.2% of the antiprotons entering the Pi2 experimental area based on TRANSPORT and GEANT4 [28] simulations). In this way, the background to signal in GAPS is $>10^5$. This challenging experimental environment is in contrast to a space-based experiment where the antiparticles have much lower energy and require only a thin degrader. Further, the goal of these experiments is not only to identify antiproton events, but also to accurately measure the individual X-ray transition yields.

The experimental setup is illustrated in figure 3. The detectors highlighted in green and labeled from above were added for the KEK05 run to provide additional information to more accurately normalize event types and rates. Since all of the particles initially have the same momentum, they can be identified by time of flight (TOF) due to their difference in mass. After entering the Pi2 experimental area, particles are tracked through the P0 and P2 plastic scintillator counters separated by 5.5 m. The timing coincidence of these counters was used as the event trigger for the detector electronics, with the TOF between providing particle identification. The extra beam counter P1 was added for the 2005 measurement to aid in identifying accidental events through TOF consistency. Only timing information was recorded for P0-P2.

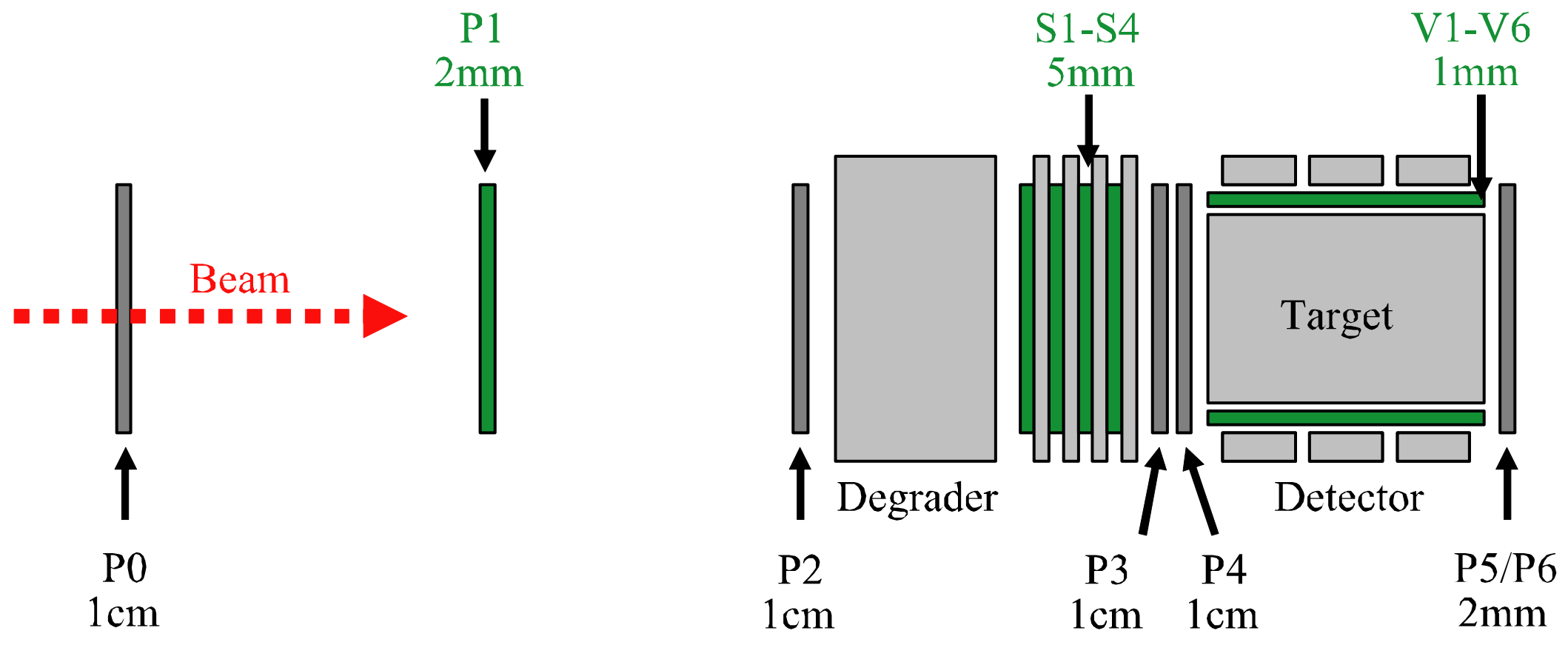

Figure 3: KEK Experimental Setup. The detectors labeled from above and highlighted in green are new for the 2005 experiment.

The P3 and P4 counters provide redundant timing and energy deposit information to tag the antiprotons which have survived passage through the degrader and enter the target. The P5 counter is used to veto antiprotons that did not stop in the target. Both timing and energy information was recorded for the P3-P5 counters. The second half of the degrader was upgraded to an active degrader for the 2005 measurement. In this way, the S1-S4 plastic scintillators interspersed with lead form a shower counter used to confirm the passage of a valid antiproton through its unique dE/dx loss signature. This significantly limits the chance of misidentifying the shower of pions and electrons (in addition to photons and neutrons) produced from direct antiproton annihilation as valid stopped antiproton events. As a further upgrade for the 2005 measurement, six charged particle veto counters (V1-V6) consisting of thin scintillating fiber bundles were added in front of the scintillator crystals to completely surround the target cell. These counters identify beam particles that scatter directly into the crystals. Energy information only was recorded for the S1-S4 shower counters and V1-V6 veto counters. In each case, the intention of the upgrades for the 2005 measurement was to provide a more reliable event tag in order to more accurately extract the X-ray transition yields and characterize background processes.

The target has a cylindrical geometry (12 cm diameter, 48 cm length) appropriate to a beamline experiment, as opposed to a flight GAPS which would employ a cubic geometry [3]. The GAPS detector and ancillary electronics are shown in figure 4a and a photograph of the GAPS detector with the gas target installed for the 2004 measurements is shown in figure 4b. Sets of 2x4 NaI crystals (25mm diameter, 5 mm thick) are housed in 16 panels arranged in a hexagonal array. Each of the 128 crystals is coupled to a Hamamatsu RM1924a photomultiplier tube (PMT). The system achieves sufficient energy resolution to resolve the X-ray transitions of interest and 200 ns time resolution for coincidence rejection of background. The level of detector segmentation was chosen to balance cost with the desire to limit the occurrence of multiple X-rays or annihilation products entering the same crystal (<3% for typical GAPS event). The solid angle coverage of the detector (35% for an average event) is limited by the open ended cylindrical geometry required for the beam entrance and exit in addition to the dead space due to NaI crystal packing inefficiencies.

A custom 128-channel data acquisition system was constructed to directly handle signals from the NaI detector phototubes. The pulse processing system is built around a gated integrator with a parallel fast channel. A discriminator in the fast channel recognizes an event and initiates the signal processing cycle. The charge from the PMT is integrated and digitized with a low power, 16 bit ADC. The system can be operated either with an external trigger (e.g., based on the TOF trigger in the KEK experiment) or in a self-triggered mode where each channel is read out every time a signal is recognized (e.g., the mode used for energy calibration of the crystals using a radioactive source). The system also digitizes 32 extra ADC channels for the timing and energy deposit of the trigger counters and records 16 logic signals for each event trigger. A fast PCI interface is used to perform high-speed data transfer between the electronics and a personal computer, and the entire system can sustain a rate of over 100 kHz per channel. The total power consumption for the data acquisition system is 150 W.

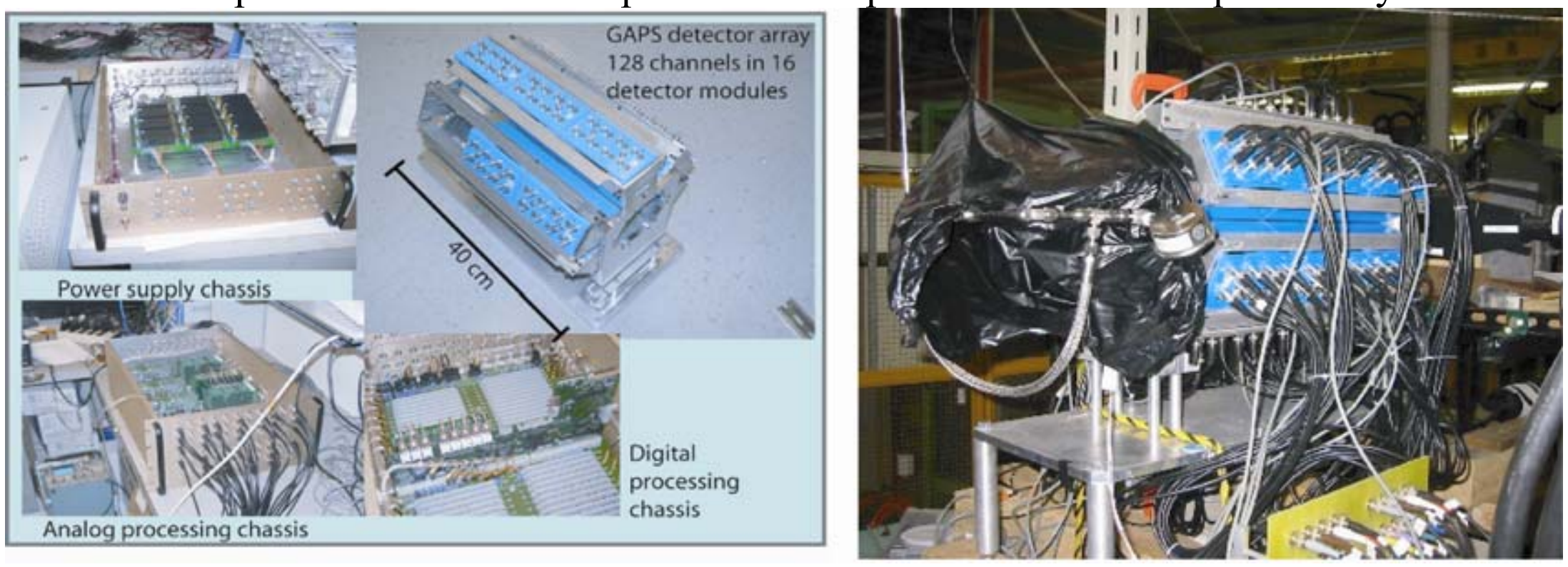

Figure 4: a) Schematic of the GAPS detector. b) Photograph of the GAPS instrument with the gas target.

## 9. Experimental Conditions

Since the rates of antiprotons are low, the beam was set up using protons ($10^5$ protons per spill). This is possible because the beamline magnets behave identically with reversed polarity. The focus at the target and the thickness of the lead degrader (~10 cm) were optimized to maximize the number of proton stops in the target (and hence stopped antiprotons as verified by GEANT simulations). The beam focus was optimized by changing an individual magnet setting in small steps over a broad range in values, measuring the proton and beam monitor signals for five seconds and setting the magnet to the value at which the maximum proton signal was obtained. Degrader range curves were performed by incrementally increasing the degrader thickness and recording the event rates in each counter. These experimental range curves were found to be in good agreement with detailed GEANT4 calculations, and the thickness for antiproton operation was set accordingly.

Event data from a +1 GeV/c run after the beamline was set up are displayed in figure 5a where the TOF between P0 and P2 is plotted versus the TOF between P0 and P1. Here, the highly relativistic pions, muons and positrons (all with $\beta \cong 1$ and hence referred to as MIP's – minimum ionizing particles) are clearly distinguished from the protons. There are approximately equal numbers protons and MIP's in each spill. A similar plot for -1 GeV/c data is plotted in figure 5b. The antiproton rate is approximately 25 per spill (the actual rate varied somewhat over the duration of our measurements due to changes in the overall rate and status of the KEK proton synchrotron). This is to be compared with a rate of ~$1.5 \times 10^5$ per spill for negatively charged MIP's. Since the antiprotons are suppressed by a factor of nearly $10^4$, our event trigger (based on P2-P0 TOF) was prescaled by a factor of 0.015 for the relativistic particles. In this way, all of the antiproton events were recorded to disc in addition to a sample of all other events for use in background studies. In each of these plots, the P2-P0 timing cut (|P2-P0| < 2 ns, highlighted by yellow) is the only timing information that was available in the 2004 measurements. A significant amount of accidental background (e.g., one particle triggering the P0 counter and different particle triggering the P2 counter) is rejected with the addition of the P1 counter, which allows for the additional P1-P0 timing cut (|P1-P0| < 2 ns, highlighted by pink).

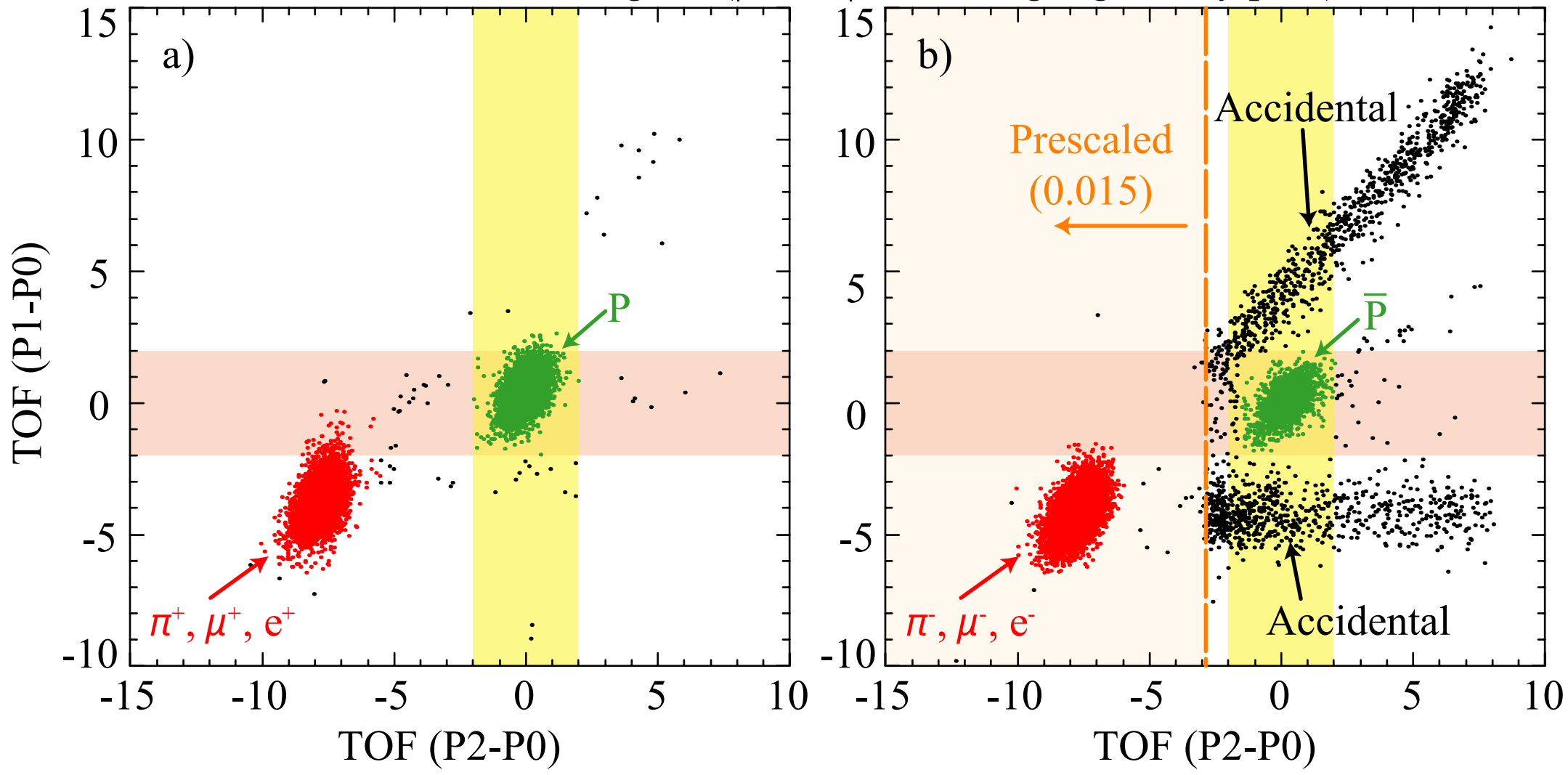

Figure 5: TOF from P0 to P1 is plotted versus TOF from P0 to P2 for +1 GeV/c particles in a) and -1 GeV/c particles in b). The protons and antiprotons are highlighted by the TOF cuts in a) and b), respectively. The P1 counter allows for the additional horizontal cut (pink) that suppresses accidental background that is present with only the P2-P0 TOF cut (vertical yellow band).

Based on the timing information, the energy deposited in the S1-S4 and P3-P5 counters was normalized using the proton range curve data (which provides particles exiting the degrader into these counters with a broad range of momentum, and hence energy deposition). While the energy resolution of the plastic scintillators in not exceptional, there is good energy separation between protons/antiprotons and the other particles, which are all essentially MIP's. Further, the redundant nature of these counters provides added assurance in making proper event identification. Even without the aid of the P1 and S1-S4 counters in KEK04, good event identification was achieved because the accidental events are all MIP's with a much smaller energy deposit (~2 MeV for MIP's instead of ~10 MeV for antiprotons) in P3 and P4 as well as a slight timing difference. The V1-V6 and P5 counters are used to veto antiproton events that exit the target without stopping. Since stopped antiparticles produce exotic atoms with nearly unity probability, the event identification power of our setup provides good normalization of the subsequent X-ray transition yields.

The following targets were used during the two runs; $C_2F_6$ (gas), $N_2$ (gas), C (aerogel), Al, S, $CCl_4$ (liquid) and $CBr_4$. The sulfur and carbon tetrabromide were powders which were put in three elliptical containers tipped at 45 degrees to the beamline. This is basically the geometry of Wiegand [22]. The aluminum was in the form of "wool" placed in a low density cylinder. The gases were held in a thin-walled (<1mm) carbon-fiber reinforced plastic (CFRP) cylinder at a pressure of 10 atmospheres. Ten atmospheres is somewhat below nominal operating pressure for a GAPS experiment [3], but high enough pressure to probe the potential effects of Stark mixing on expected X-ray yield. The CFRP pressure vessel was designed to operate (and successfully tested) at 40 atmospheres, but KEK safety procedures precluded operation at that pressure in these runs. Typical target thicknesses were ~7 g/cm$^2$, chosen to provide good stopping power and X-ray transparency. In this way, antiprotons entering the target with kinetic energy less than 100 MeV would be captured in GAPS.

The primary experiments were performed with beam momentum of -1 GeV/c. In addition to antiprotons, data runs were made using a proton beam and negative pion beams with energies between 0.2 and 1.2 GeV since these particles are a background source in space-based experiments. Runs were also made with antiprotons at -1 GeV/c in configurations with GAPS but no degrader and with a degrader but no target in GAPS. This data will allow precise characterization of background event topologies in GAPS for comparison with simulations. Below we present some very preliminary analysis which demonstrates, for the first time, the simultaneous detection of multiple ladder transition X-rays and pion stars, thus verifying the basic GAPS concept. A much fuller analysis will be presented in subsequent papers. However the results of even this preliminary analysis confirm the basic assumption about GAPS performance, which were based on previous sensitivity calculations [3].

Antiprotons were successfully identified in an environment with S/B of ~$10^{-5}$. The detectors external to GAPS comprised the shower, timing and scintillating fiber veto counters shown in figure 3. These detectors provide a detailed means to identify each particle entering and/or stopping in GAPS with extremely high fidelity. However it is crucial to note that our off-line analysis has enabled us to identify each stopped antiproton directly through its unique X-ray and pion star signature, with no recourse to these external detectors.

## 10. Experimental Results

A variety of solid, liquid and gas targets were used in KEK04 and KEK05, as described above. Since the goal of this paper is simply to present some preliminary data indicating that the basic concept is sound, we restrict our discussion to just a few of these targets. More importantly, the primary goal was to evaluate candidate targets for real space experiments. In that regard high molecular weight organic gases are superior to nitrogen – our original baseline target - because of better stopping power and higher effective X-ray yield. Carbon aerogel is superior to gases (for engineering design reasons), and all the solid and liquid targets of KEK05 are superior to Carbon aerogel (because the 3rd X-ray transition of interest is at higher energy and thus is detected with higher efficiency). So the experimental program has already led to an evolution of the GAPS concept away from gases.

During the time in which we ran with an ethyl hexafluoride ($C_2F_6$) target, approximately 1500 exotic atoms were formed. A histogram of all X-rays detected in this target is plotted in figure 6a. The X-ray spectrum with the target removed is also plotted (1.5 hours equivalent data scaled to the 31 hours of ethyl hexafluoride data). Based on the ambient X-ray background spectrum and total rate (~10 X-rays/sec/crystal), the total accidental X-rays in the triggered antiproton data is ~100 X-rays. The data set can be corrected for the number of accidentals in each X-ray line of interest. Both GEANT4 and separate Monte-Carlo simulations were used to determine the expected count rate in the X-ray lines. These calculations take into account the triggered antiproton rate on target, the efficiency of detector and electronics and an assumed effective X-ray yield. We expect to see somewhere on the order of 100 X-rays in each exotic transition. This is a preliminary estimate and is comparable to observations. More detailed analysis is ongoing. A comparable calculation to this was performed for each of the targets. The effective X-ray yields, even based on preliminary calculations with a factor of ~2 uncertainty, are consistent with the assumption about X-ray yield in [3]. This is to be expected. While the antiprotonic X-ray yield has only been measured in a limited number of elements (such as nitrogen, the original GAPS baseline), the antiprotonic yields are expected to be comparable to those measured in kaonic exotic atoms, where yields have been measured for all elements and many compounds. We selected targets all known to have kaonic X-ray yields in excess of 30%, some even larger than 50%. Analysis is underway to directly compare the X-ray yield of our targets as a function of Z with Fermi-Teller theory, as has been done extensively for kaonic atoms.

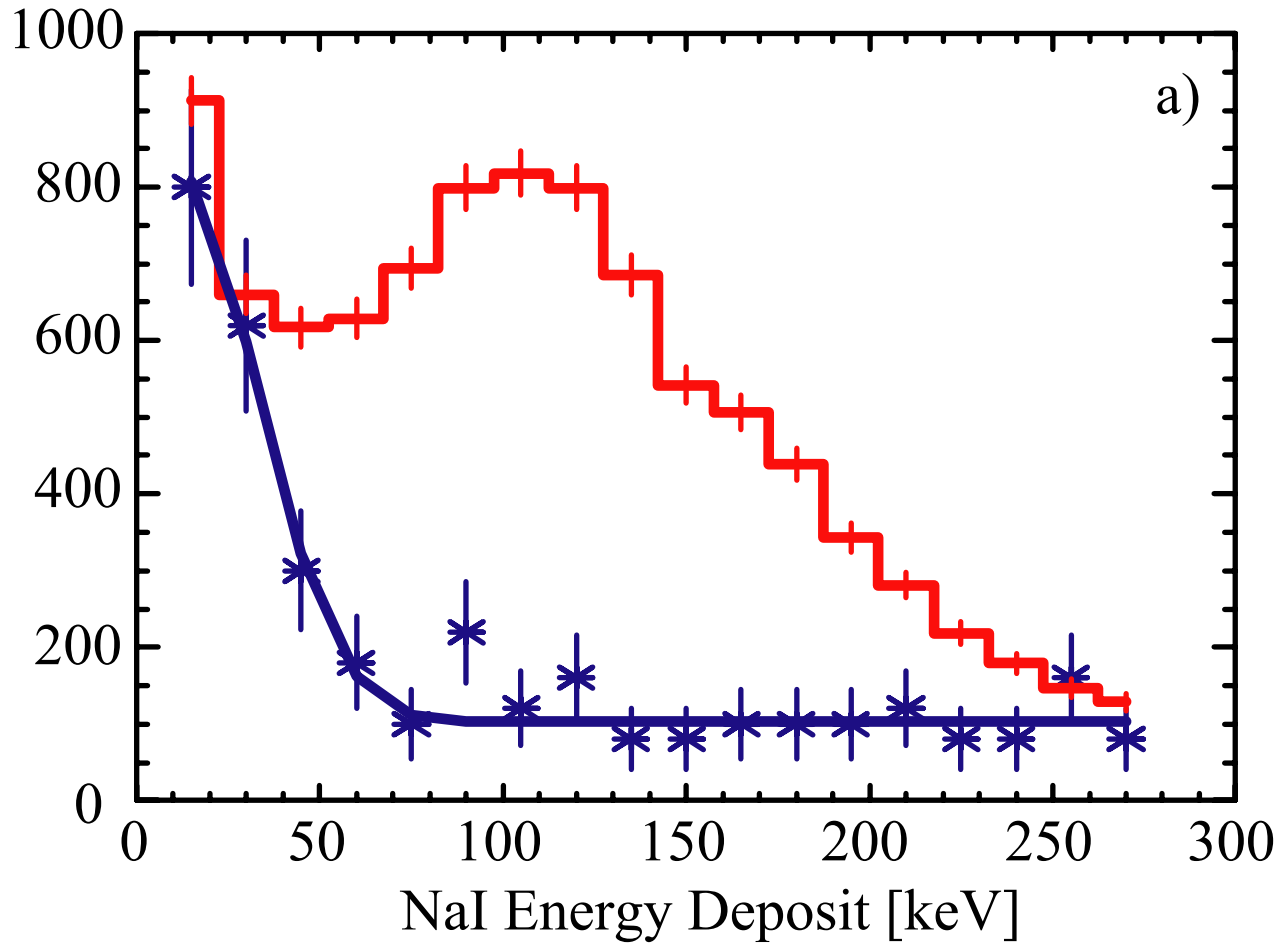

Figure 6: a) Histogram of X-rays for $C_2F_6$ target. The X-ray spectrum with the target removed is also plotted. b) GAPS event topology in the $C_2F_6$ gas target for a 3 X-ray and a charged pion star tag (π*).

In addition to these transition X-rays, we expect to see other energy deposits associated with annihilation of the antiproton in the nucleus, as described earlier. Charged and neutral pions are emitted during the annihilation. The charged pions leave an enormous amount of energy as they pass through the NaI crystal (~ 2 MeV) and are recorded as a pion star signal (π*). Because the nominal X-ray ladder transitions in ethyl hexafluoride are well-below 100 keV, resolving the X-ray lines in the integrated spectrum is not possible. However on an event by event basis the GAPS event topology provides a clear indication that an antiproton has stopped in the gas. Figure 6b shows an antiproton stop which produces three X-rays at the proper energy along with a charged pion, all in time coincidence. This is a signature of an antiproton annihilation. Here, as in all the events presented below, the experiment trigger as defined in section 9 independently confirms that the event in question really is an antiproton, as the topology indicates.

Accurate measurement of X-ray transition energies is important for identifying valid antiproton events. During KEK04 X-ray calibrations with check sources (americium and barium) were run every hour over the entire NaI array. This data was used to correct for temperature drifts (diurnal temperature changes in the experimental hall were as large as 25C). Temperature conditions were more stable in KEK05 requiring less frequent X-ray calibrations.

The KEK05 run emphasized solid targets, based on the promising results obtained with Carbon aerogel in KEK04. The most interesting target is carbon tetrabromide ($CBr_4$). Carbon tetrabromide produces up to seven X-ray ladder transitions within the high detection efficiency bandwidth of a thin NaI crystal, many at higher energies. The background rejection power of GAPS goes like $(\Delta E \Delta \tau)^n$, where $\Delta E$ is the energy resolution, $\Delta \tau$ the system time resolution and n the number of detected X-rays. The minimum required for antideuteron searches is n=3 (c.f., [3], where typically n=3 for gases).

Moreover the sensitivity of GAPS is dominated by the lowest energy X-ray, which has the lowest probability of escaping to the target. Thus X-rays of higher energy are most beneficial for GAPS. All the solid and liquid targets investigated can provide three or more X-rays where the lowest X-ray has higher energy than in any of the gases considered in [3] or the targets of KEK04.

Figure 7a shows a carbon tetrabromide integrated X-ray spectrum. This spectrum was produced solely using cut criteria intrinsic to GAPS. In this case the cut required more than two ladder X-ray transitions and more than four total energy deposits. The spectrum of figure 7a clearly shows the bromine X-rays (whose capture probability is large compared to the carbon). Again correcting for accidentals, the X-ray ladder transition rates are consistent with high effective X-ray yield – comparable to those in the corresponding kaonic system. More detailed analysis is underway. There are several unidentified lines above ~170 keV in bromine and other solid target spectra. Preliminary analysis suggests these are the nuclear excitation lines associated with the daughter nuclei produced in antiproton annihilation of the nucleus, as first observed by Barnes [29]. We were unaware of the existence of such lines until the KEK experiments, and did not consider their potential salutary impact on the sensitivity and particle identification capability of GAPS in our original analysis. These lines, at well-defined energies, can serve to augment the atomic X-ray ladder transitions. Because these nuclear lines are all at higher energies, they are particularly useful because they easily escape the target.

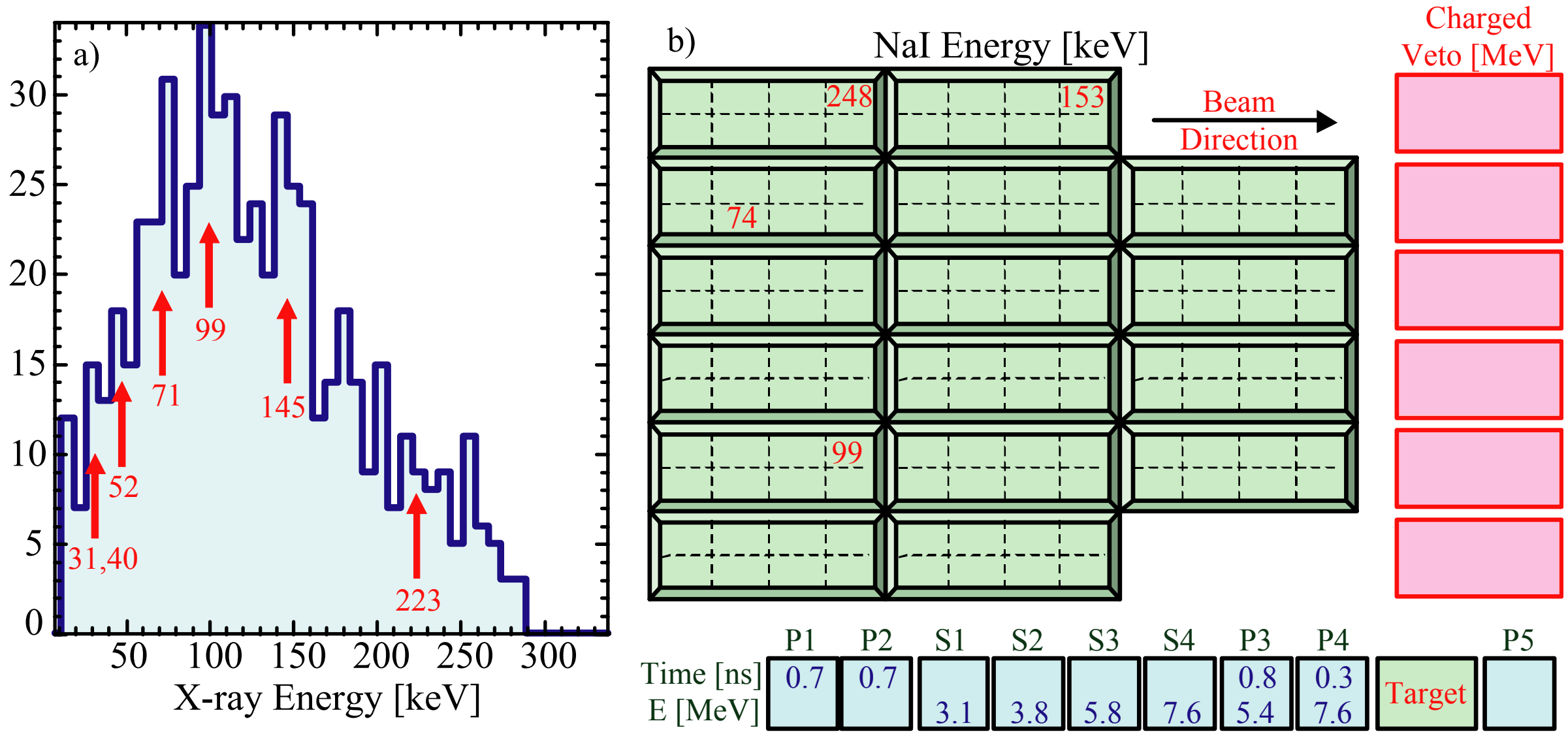

Figure 7: a) The carbon tetrabromide ($CBr_4$) integrated X-ray spectrum for events with ≥2 ladder X-ray transitions and ≥4 total energy deposits. b) An example of a four ladder X-ray event.

Figure 7b shows the signature of a carbon tetrabromide antiproton annihilation. Four ladder X-rays are cleanly identified but there are no associated nuclear annihilation pions. Given that the mean number of pions per annihilation is approximately five, this at first seems striking. However the solid angle coverage of our prototype GAPS is rather modest at ~0.3. In addition any neutral pions will decay immediately to gamma-rays with very small probability of energy deposit in our thin crystals. Thus the existence of such events is expected in the prototype. Figure 7b also shows how we are able to confirm, independently of the GAPS event signature, that the event was induced by an antiproton. The boxes surrounding the segmented crystal display give data on the upstream and downstream diagnostics. In particular, displays P1-P4 for the plastic scintillators indicate the timing deviation at each piece of plastic compared to that expected for an antiproton. In each case the timing deviation is within the timing resolution of the system (~1 ns). The counters S1-S4 show the monotonic increase in deposited energy whose magnitude in each counter is consistent with that expected for slowing down of an antiproton of proper incident momentum. And finally there is no signal in the downstream plastic P5 (indicating a stop in GAPS) and no signal in the scintillating fiber veto inside GAPS which would indicate an antiproton which elastically scattered out of the target cell. Elastically scattered antiprotons can be identified by the signatures they produce on capture and annihilation in the NaI itself.

The integrated spectrum for a sulfur target is shown in figure 8a using the same energy cuts as in figure 7a. Figure 8b shows an event where two ladder X-rays are detected along with three annihilation pions. These events are of particular interest to us in our current analysis. The original analysis of Mori et al. [3], while recognizing the existence of the annihilation pions, did not explicitly consider them in calculating the particle identification capability of GAPS. However it is quite clear that when combined with a few ladder X-rays, the pions are an extremely powerful additional means to identify antiparticle stops. In fact the pions, because their stopping power allows them to penetrate several detection cells of the cubic geometry, provide a tremendous opportunity to use GAPS as a tracking/timing chamber to help identify antiparticles. The original GAPS analysis of flight configurations did consider the potential value of higher energy ladder X-rays which Compton scatter in one target cell and then are photoelectrically absorbed in another. But the pions are likely to be an even more powerful tool for enhancing the X-ray signature or making up for the loss of a ladder X-ray. Note that unlike the event of figure 7b, the event in figure 8b has an associated signal in the charged particle veto. The magnitude of the energy deposits is indicative of an annihilation pion transit, not antiproton scatter in the target.

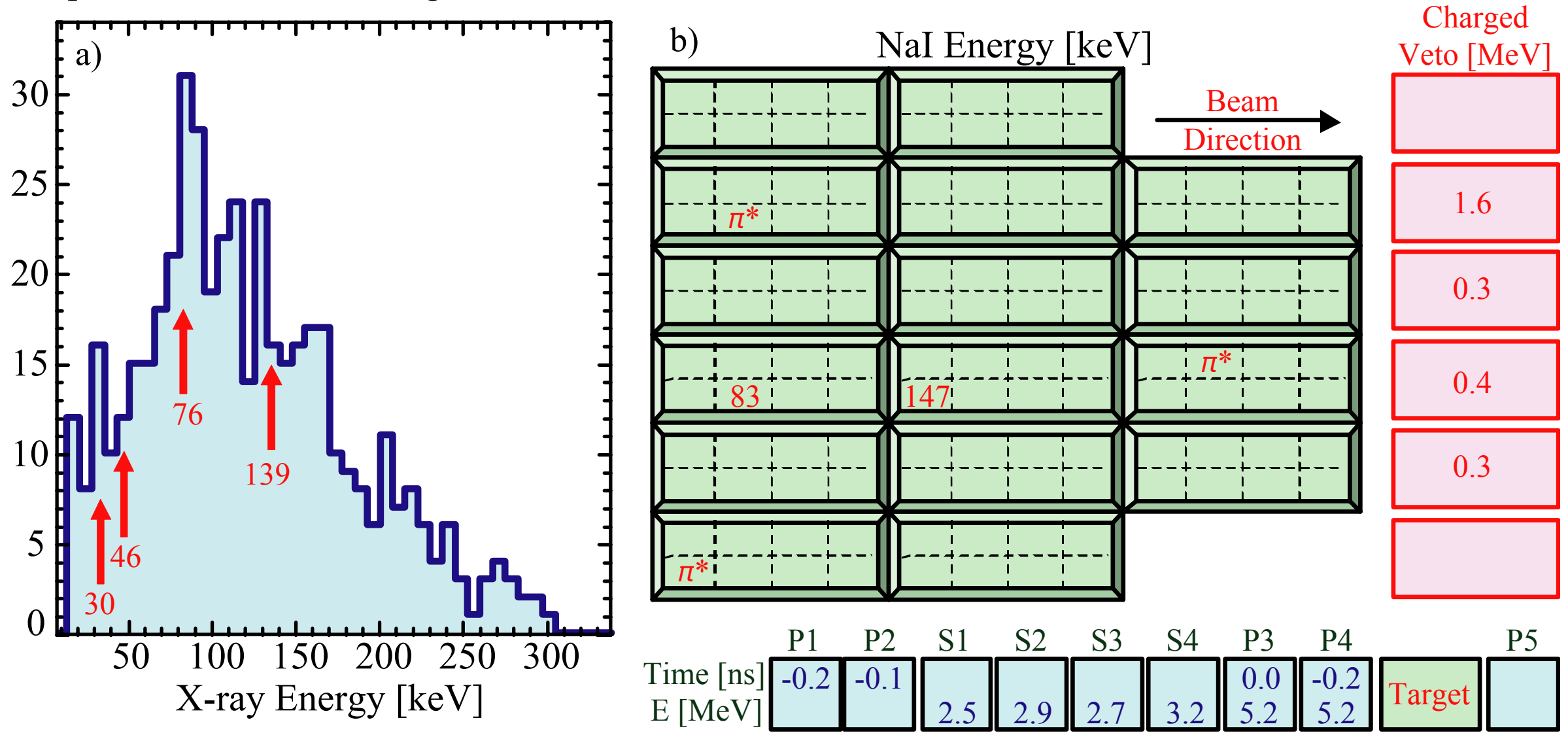

Figure 8: a) The Sulfur integrated X-ray spectrum for events with ≥2 ladder X-ray transitions and ≥4 total energy deposits. b) An example of a two ladder X-ray event.

As a final example of the type of data obtained in the KEK runs, figure 9 shows an event whose topology is consistent with being an antiproton scattered in the GAPS detector. No energy deposit is consistent with a ladder transition, and the topology of a track moving through multiple crystals is present. This indicates a non-annihilating, charged particle. This is confirmed independently by upstream and downstream counters which indicate an antiproton entered GAPS and was scattered into the NaI array (evidently at an oblique angle). All the scattered antiprotons have readily identifiable topologies.

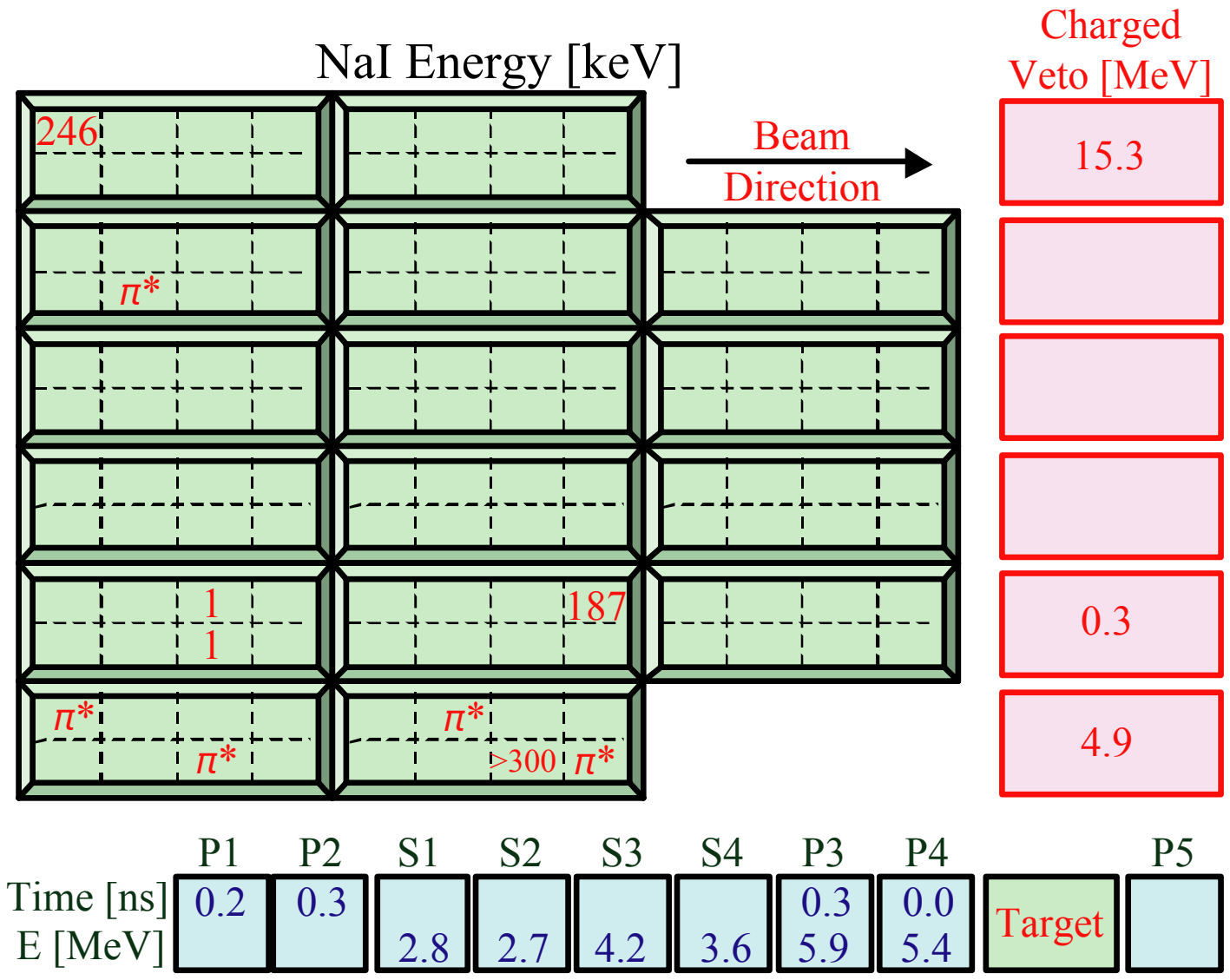

Figure 9: Antiproton scattered into GAPS detector.

Although not presented here, considerable data was also taken with protons, pions and muons to provide basic data to assist in estimating the probability of particle misidentification. For example, we reversed the polarity of the beam magnets and recorded approximately the same number of proton triggers as antiprotons for each target. In the ethyl hexafluoride data set, for instance, the X-ray event rate is significantly less than for antiprotons, and no multiple X-ray events were ever produced, much less a three or four signature event. This is consistent with computer simulations. Much more on this will be reported elsewhere. Another cross-check was to add 2 cm of extra lead to the degrader to stop all of the antiprotons before they enter the target, but not other beam related background. No multiple-signature events were recorded in this data set. These results, while comforting, are not unexpected. Previous modeling [3] suggests that the primary source of particle misidentification in GAPS will be caused by simultaneous energy deposits (X-rays and beta particles) resulting from cosmic-ray activation of the detector and surrounding material (in coincidence with a TOF trigger). Nevertheless it is reassuring that preliminary analysis shows that the particle background is indeed unable to produce a problematic false signature.

## 11. Experimental Implications and Future Directions for GAPS

The most important result of the KEK04 and KEK05 runs was to verify the basic GAPS concept. Simultaneous atomic ladder X-ray transitions and pion stars have been used, for the first time, to specifically identify antiparticles. GAPS provided near unity efficiency in identifying antiproton stops. These results were confirmed with a completely independent array of experiment detectors which identify particles, by type, which stop in GAPS.

Two results of great importance to future design work were obtained. Firstly, solid (and liquid) targets have been successfully utilized. This enormously simplifies the design challenges of GAPS. With no need for high pressure gas, as in the original concept, GAPS is easier to operate and it is lighter and more efficient because of the removal of the dead mass of the gas handling system. GAPS efficiency also increases because solid and liquid targets provide more design options. Gas targets must be both transparent to their ladder X-rays and immune to Stark mixing at high gas pressure. These were restrictive conditions given the limited number of gas candidates. Solid and liquid targets permit more choices where there are both more ladder X-rays in the detector bandpass and X-rays of higher mean energy. This translates directly into higher sensitivity, since loss of X-rays and particularly the lowest energy X-ray (the hardest to get out of the target and into the detector) dominates the sensitivity of GAPS. And secondly, the preliminary estimate of effective X-ray yield gives numbers consistent with the >~30% used in the original GAPS sensitivity calculations. Our conclusion from the experiments is that the GAPS concept is sound and that the sensitivity numbers of Mori et al. [3] are basically correct. Thus the theoretical implications of a GAPS experiment, discussed in numerous papers, will not be meaningfully altered when exact effective X-ray yields are produced from our detailed analysis. Two other results are of potential importance, although much more work will be required to understand their implications. Firstly the pion stars provide substantial additional antiparticle identification capability, ignored in the original GAPS work. This added capability is likely to be even more important in a real space experiment, since the cubic array geometries proposed for space-based GAPS are ideally suited to exploiting the track which these pions will produce on escaping the GAPS cell in which they were created. Secondly the nuclear deexcitation lines, correlated with the atomic ladder X-rays, are potentially a source of added confirmation that an antiparticle stop has taken place.

The experimental program at KEK in Japan has been sufficiently successful to move on to a flight test of GAPS. Our current plan is to use the KEK GAPS prototype. The prototype cylindrical, single cell, geometry is unlike that proposed for a flight GAPS in [3]. Nevertheless it will be extremely useful for evaluation of in-flight background, since a key GAPS parameter is how effectively it can reject the copious charged particle and cosmic-ray activation background. A successful experiment would be one where all this background is identified and rejected. The plan would be for a 24 hour flight from Lynn Lake, Canada. We note that the electronics used at KEK were flight representative. GAPS could be prepared for such a flight in 3 years or less.

Our expectation is that such a test flight would take place in parallel with work to evaluate advanced flight detector and readout concepts. While NaI is a leading candidate for the X-ray detector, there are other promising approaches. CZT was specifically mentioned in our original GAPS work. Another option is CsI(pure), which reduces mass because of its superior stopping power compared to NaI(Tl) and is also cheaper. CsI light output is less than NaI(Tl) but it compensates for this by having better time response. New materials such as LaCl offer improved energy resolution albeit with higher cost. Perhaps the highest leverage for simplifying and improving on GAPS is in the readout for crystal detectors. A crystal bar geometry with miniature PMTs on either end (such as those available in T0-5 cans from Hamamatsu) is a more natural readout geometry than in the GAPS prototype, where the conventional crystal arraying leads to large dead area. Of particular utility to GAPS may be to explore using waveshifting materials to convert NaI(Tl) optical emission to the green region, where avalanche photodiodes or PMTs provide excellent quantum efficiency. This approach, which was used by Ziock and Hailey [30], employs wavelength-shifting fibers abutted to the scintillator bars to detect the bluish scintillation light and downshift it to green for efficient transport in the fiber. The light is detected in a position sensitive PMT, and the pixel of the PMT which is hit gives the “address” of the detector cell. Fiber runs of about a meter are feasible with the light yields expected from ladder X-rays, and this type of system has been demonstrated to work exceptionally well [31]. This approach has a large multiplex advantage though this must be traded against energy resolution. In this scheme virtually all electronics are outside the detector cells, minimizing dead mass.

While much work remains to be done, it is realistic to envision flying a GAPS instrument that can detect antideuterons with sufficient sensitivity to probe well into the CMSSM and other SUSY parameter spaces on a timescale of five years or less. The precise timescale will be driven, as always, by funding.

**Acknowledgements**
We thank J Collins and the electronics shop staff at LLNL for the development and construction of the GAPS electronics, and T Decker, R Hill and G Tajiri for mechanical engineering support. We would also like to thank J Jou for his assistance during the 2005 KEK beamtime. We gratefully acknowledge the support of M Ieiri and the KEK staff before and during the accelerator experiments. This work was supported in part by a NASA SR&T grant, NAG5-5393.